\documentclass[12pt]{article}
\usepackage{amsmath}
\usepackage{amssymb}
\usepackage{graphicx}

\input{epsf}

\begin{document}

\begin{center}
{ \large \bf 
Quantum automata, braid group  
and link polynomials}
\end{center}

\vspace{12pt}

\noindent 
{\large {\sl Silvano Garnerone}}\\
\noindent Dipartimento di Fisica,
Politecnico di Torino,\\
corso Duca degli Abruzzi 24, 10129 Torino (Italy)\\ 
E-mail: silvano.garnerone@polito.it \\

\noindent
{\large {\sl Annalisa Marzuoli}}\\
\noindent Dipartimento di Fisica Nucleare e Teorica,
Universit\`a degli Studi di Pavia and 
Istituto Nazionale di Fisica Nucleare, Sezione di Pavia,\\ 
via A. Bassi 6, 27100 Pavia (Italy)\\ 
E-mail: annalisa.marzuoli@pv.infn.it \\

\noindent
{\large {\sl Mario Rasetti}}\\
\noindent Dipartimento di Fisica, 
Politecnico di Torino,\\
corso Duca degli Abruzzi 24, 10129 Torino (Italy)\\ 
E-mail: mario.rasetti@polito.it \\

\
\begin{abstract}
\noindent The spin--network quantum simulator model, which essentially encodes 
the (quantum deformed) $SU(2)$ Racah--Wigner tensor algebra, 
is particularly suitable to address problems arising in 
low dimensional topology and group theory.
In this combinatorial framework we implement families of
finite--states and  discrete--time quantum automata 
capable of accepting the language generated by the braid group,
and whose transition amplitudes are colored Jones polynomials.
The automaton calculation of the polynomial of
(the plat closure of) a link $L$ on $2N$ strands
at any fixed root of unity is shown
to be bounded from above by a linear function of the number of
crossings of the link, on the one hand, and polynomially
bounded in terms of the braid index $2N$, on the other.
The growth rate of the time complexity function in terms
of the integer $k$ appearing in the root of unity $q$ can be
estimated to be (polynomially) bounded by resorting
to the field theoretical background given by the Chern--Simons 
theory.
\end{abstract}


\vfill
\newpage

\section{Introduction}

The spin--network quantum simulator model \cite{MaRa1,MaRa2} represents a bridge between
circuit schemes for standard quantum computation
and approaches based on notions from Topological 
Quantum Field Theories (TQFT) \cite{FrKiWa1,FrKiWa2,FrKiLaWa}. 
The spin--network computational space, naturally modelled 
as a graph for any fixed number of incoming spins, supports computing
processes represented by families of paths and
provides, on the one hand, a consistent
discretized version of the topological quantum computation approach.
On the other hand, such a quantum combinatorial scheme, which essentially encodes 
the (quantum deformed) $SU(2)$ Racah--Wigner tensor algebra, 
turns out to be particularly suitable to address problems arising in (low
dimensional) topology and group theory.
The guiding idea of this paper is  that the exponential efficiency that quantum algorithms may achieve 
with respect to classical ones proves to be especially relevant in problems in which the space of 
solutions is characterized by a structure definable in terms of the grammar and the syntax of a language, 
rather than  algebraic or number--theoretic in nature. 
The spin--network  setting provides a  
`natural encoding' for classes
of problems  which basically share the combinatorial structure of the language underlying the 
(re)coupling theory of $SU(2)$ angular momenta \cite{BiLo9}.

On the other hand, the Jones polynomial \cite{VJo} is no doubt the most famous knot invariant in topology, 
a knot invariant being a 
function on knots (or links, namely circles embedded in $3$--space) which 
is invariant under isotopy (smooth deformations) of the knot. 
Among its many connections to various mathematical and physical areas (see {\em e.g.}
\cite{Wu} for applications in statistical mechanics), we are mainly interested here in its 
relations with TQFT \cite{Ati}.
In the seminal paper \cite{Wit}, Witten put link invariants
in a field theoretical setting, showing that Jones polynomials 
arise as  vacuum expectation values of 
Wilson loop operators in a three dimensional $SU(2)$ Chern--Simons (topological) quantum 
field theory where the fundamental 
representation of the gauge group $SU(2)$ lives on each component of the link. Such an invariant
was extended  to arbitrary representations living on the link components and 
in this paper we shall deal with such generalizations, referred to as `extended' or
`colored' Jones polynomials \cite{ReTu,KiMe}.

From the (classical) computational side, it was proved that the exact evaluation of the
Jones polynomial of a link $L$, $V(L,\omega)$ at $\omega =$ root of unity, can be performed in
polynomial time in terms of the number of crossings of the planar diagram of
$L$ if $\omega$ is a 2nd, 3rd, 4th, 6th root of unity. Otherwise, the problem is 
$\# {\bf P}$--hard \cite{JaVeWe} (the computational complexity class 
$\# {\bf P}$--hard is the enumerative analog of the ${\bf NP}$ class).
However, Kitaev, 
Larsen, Freedman and Wang \cite{FrKiWa2} showed that their `topological' quantum
computation setting, relying on the same TQFT  quoted before, 
implicitly provides  an efficient quantum algorithm for the approximation 
of the Jones polynomial at  a fifth root of unity. Unfortunately, this important algorithm was never 
explicitly formulated. This is particularly unfortunate since it is known that the approximation problem is 
{\bf BQP}--hard, and a quantum algorithm for this problem is thus of particular importance.

Let us  point out that recently Aharonov, Jones and Landau  proposed an efficient quantum algorithm 
that approximates the problem of evaluating the Jones polynomial based, rather than on physical 
results from TQFT, on the path model representation of the braid group and the uniqueness of the Markov 
trace for the Temperley--Lieb algebra \cite{AhJoLa}. The argument is that the $\# {\bf P}$--hardness 
of the problem does not rule out the 
possibility of good approximations, and indeed these authors provide an efficient, 
explicit and simple quantum algorithm to 
approximate the Jones polynomial at all roots of unity for both the trace and the plat 
closures of a braid.\\ 
Our strategy is quite different from theirs, since we shall
basically provide a quantum (automaton) system whose
internal evolution can be controlled in such a way that its
probability amplitude gives the desired polynomial.

\begin{figure}[htbp]
\begin{center}
\includegraphics[height=6cm]{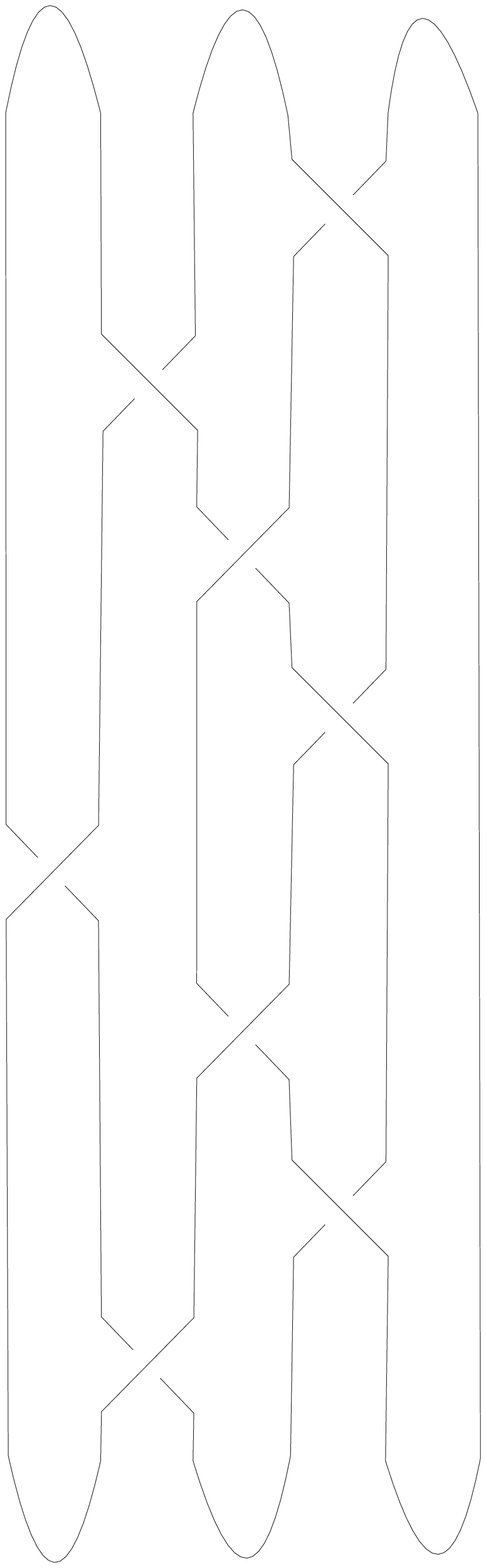}
\end{center}
\caption{A plat presentation of the borromean link.}
\label{borromeanplat}
\end{figure}

As mentioned, one of the features of the Jones polynomial that will be
used extensively is that it can also be defined 
via braids (a geometric $N$--braid is a set of $N$ strands with fixed endpoints in the plane).
A braid can be `closed up' to form a link by tying its ends together. In this paper we shall be 
interested in one of the two ways to perform such closures, namely the plat closure of the braid,
 and hence consider extended Jones polynomials associated with such 
link diagrams, {\em cfr.} Fig. \ref{borromeanplat}. 

On a broader front, the study of braid groups and their applications is a field which has attracted  
great interest from  physicists, mathematicians and computer scientists alike
({\em cfr.} \cite{BiBr} for an updated review). Besides for its value in 
studying the braids in a theoretical framework, 
applications to knot theory have been known for 
years, while applicability to the field of cryptography has been realized recently
\cite{AnAn}. 
The analysis of algorithmic 
problems related to braid group has thus acquired 
a great practical significance, in 
addition to its intrinsic theoretical interest.  

The approach we present here exploits
a $q$--braided version of the original spin--network setting
\cite{MaRa2}  to make it accept the language of
the braid group and to deal with 
link polynomials
(see \cite{GaMaRa} for a presentation of some preliminary results). 
As pointed out before,
the `physical' background provided by the 3$D$ quantum $SU(2)$
Chern--Simons field theory  plays a prominent role,
because our computational scheme is actually designed
as a discretized conterpart of  the topological quantum
computation setting proposed in \cite{FrKiLaWa}. Moreover, 
this framework is exactly what  is needed  to deal
with (normalized) $SU(2)$--colored link polynomials 
expressed as vacuuum expectation values of composite 
Wilson loop operators, on the one 
hand, and with unitary representations
of the braid group, on the other.
These expectation values, in turn, will provide a bridge between the theory of formal languages 
and quantum computation, once more having as natural  arena for discussion the 
$q$--braided spin--network 
environment.
 We are going to implement families of
finite states (and discrete time)--quantum automata 
capable of accepting the language generated by the braid group,
and whose transition amplitudes are colored Jones polynomials.
More precisely, our results will be interpreted in terms of 
`processing of words' --written in the alphabet given by the generators of the braid 
group-- on a quantum automaton in such a way 
that the expectation value associated with the internal automaton
`evolution' is exactly 
the extended Jones polynomial. 
The quantum automaton in question will in turn 
correspond to a path in the $q$--braided spin--network computational graph. 
The calculation of the polynomial of
(the plat closure of) a link $L$ on $2N$ strands will be shown
to be bounded from above by a linear function of the number of
crossings of the link, on the one hand, and polynomially
bounded in terms of the braid index $2N$, on the other.
Notice that the growth rate of the time complexity function in terms
of the integer $k$ appearing in the root of unity $q$ can be
easily estimated to be (polynomially) bounded by resorting
to the TQFT background, since $k$ is nothing but the Chern--Simons 
coupling constant.

We shall leave as open problems the analysis of the complexity of
the preparation of (initial and final)  states as well as the efficient 
implementation of the 
individual automaton transition functions, which might be addressed by means of
approximating (classical or quantum) algorithms.\\
In conclusion, we argue that our field theoretical approach could be further generalized,
by suitable modifications of the braiding prescriptions
in the spin--network scheme, 
to deal with 2--variables link polynomials such as the HOMFLY invariant \cite{Homfly},
related to the partition function of Potts model \cite{Wu}.
 
\vspace{12pt}

The content of the paper is, as far as possible, self contained.
In section 2 we briefly recall the definitions of classical and 
quantum languages and finite states--automata. In section 3 
we give a review of the spin--network computational framework
modelled on the Racah--Wigner tensor algebra of $SU(2)$. 
In section 4 we  deal with the $q$--braided version of
the spin--network simulator, which relies on the tensor
algebra of $U_q(su(2))$ (at $q=$ root of unity).
Section 5 is splitted into two parts: in 5.1 we review
the `quantum group approach' (and related $R$-matrix)
to the study of (unitary) braid group
representations and `quantum' link invariants; in 5.2 we
present the field--theoretical background (Chern--Simons TQFT,
Wess--Zumino boundary theory, composite Wilson loop operators
and their expectation values) trying to resort to geometric
intuition rather than to a deep knowledge of techniques in quantum field theory.
In section 6 we explain in details the automaton calculation of
the extended Jones polynomial.

\section{Classical and quantum formal languages}

The theory of automata and formal languages addresses in a rigorous way the notions 
of computing machines and computational processes. 
If $A$ is an alphabet, made of letters, digits or other symbols, and $A^*$ denotes
the set of all finite sequences of words over $A$, a language 
${\cal L}$ over $A$ is a subset of $A^*$. 
 The length of the word $w$ is denoted by $|w|$ and $w_i$ is its $i$'th symbol. The empty word 
is $\emptyset$ and the concatenation of two words $u$, $v$ is denoted simply  by $uv$. In the sixties Noam 
Chomsky introduced a four level--hierarchy describing formal languages according to their structure
(grammar and syntax): regular languages, 
context--free languages, context--sensitive languages and recursively enumerable languages.
The processing of each language is inherently related to  a particular computing model
(see {\em e.g.} \cite{HoUl} for an account on 
formal languages). 
Here we are interested in finite states--automata, 
the machines able to accept regular languages. 

A deterministic finite state automaton (DFA) consists of a finite set of states $S$, an input alphabet $A$, 
a transition function $F:S\times A\rightarrow S$, an initial state $s_{in}$ and a set of accepted states 
$S_{acc}\subset S$. The automaton starts in $s_{in}$ and reads an input word $w$ from left to right. At 
the $i$--th step, if the automaton reads the word $w_i$, then it updates its state to $s'=F(s,w_i)$, where $s$ is the 
state of the automaton reading $w_i$. One says that the word has been accepted if the final state reached after 
reading $w$ is in $S_{acc}$.\par
\noindent In the case of a  non--deterministic finite state automaton (NFA), the transition function is defined 
as a map $F:S\times A\rightarrow P(S)$, where $P(S)$ is the power set of $S$. After reading a particular 
symbol, the transition can lead to different states, according to some assigned probability distribution . 
If a NFA has $n$ states, 
for each symbol $a\in A$ there is an $n\times n$ transition matrix $M_a$ for which $(M_a)_{ij}=1$ if and only if the 
transition from the state $i$ to the state $j$ is allowed once  the symbol $a$ has been read.
 
Generally speaking,
quantum finite states--automata (QFA) are obtained from their classical probabilistic counterparts by moving 
from the notion of (classical) probability associated with transitions to 
quantum probability amplitudes. Computation takes place inside 
a suitable Hilbert space through unitary matrices and a number of different models have been proposed, see
{\em e.g.} \cite{AmWa,MoCr},
just to mention a couple of them.  
Following \cite{MoCr}, the measure--once
quantum automaton is a 5-tuple $M=(Q,\Sigma,\delta,\mathbf{q}_0,\mathbf{q}_f)$, where $Q$ is a finite set 
of states, $\Sigma$ is 
a finite input alphabet with an end--marker symbol $\#$ and $\delta:Q\times \Sigma\rightarrow Q$ is the transition 
function. Here $\delta(\mathbf{q},\sigma,\mathbf{q}')$ 
is the probability amplitude for the transition from the state $\mathbf{q}$ to the state 
$\mathbf{q}'$ upon reading the symbol $\sigma$. 
The state $\mathbf{q}_0$ is the initial configuration of the system, and $\mathbf{q}_f$ 
is an accepted final states. For all 
states and symbols the function $\delta$ must be unitary. The end--marker $\#$ is the last symbol of each input 
and computation terminates after reading it. At the end of the computation the automaton measures its configuration: 
if it is an accepted state then the input is accepted, otherwise is rejected. The configuration of the automaton 
is in general a superposition of states in the Hilbert space where the automaton lives. The transition function is 
represented by a set of unitary matrices $U_{\sigma}(\sigma\in\Sigma)$, where $U_{\sigma}$ represents the unitary 
transition of the automaton reading the symbol $\sigma$. The probability amplitude for the automaton of accepting 
the string $w$ is given by
\begin{equation}\label{autamp}
f_M \left( w \right) = \left\langle \mathbf{q}_f  \right|U_w \left| \mathbf{q}_0  \right\rangle, 
\end{equation}
and the explicit form of $f_M (w)$  defines the language $\mathcal{L}$ accepted by that particular automaton.
If $\hat{P}$ denotes the projector over the accepted states, the probability 
for the automaton of accepting the string $w$ is given by 
\begin{equation}\label{autprob}
p_M (w) =  \lVert  \hat{P} \,|\psi _w  \rangle  \rVert ^2 
\end{equation}
where $|\psi_w \rangle  \doteq U_w\,| \mathbf{q}_0 \rangle$.

\section{The quantum spin--network simulator}
The spin network model of computation was introduced in \cite{MaRa1} and worked out in \cite{MaRa2}
as a general framework for processing information
in the quantum context and is essentially modelled on the combinatorics of the Racah--Wigner algebra
of $SU(2)$.
The spin--network can be seen as a collection of graphs $\mathfrak{G}_n(V,E)$ parametrized by an integer 
$n$ ($n\geq 2$), 
where $n+1$ is the number of incoming angular momentum variables, each associated with an irreducible
representation (irrep) of $SU(2)$, $\{j_i\} \in$ $\{0,1/2,1,3/2\ldots\}$ in $\hbar$ units (we choose
units in which $\hbar=1$). 
On the physical side, these $n+1$ basic variables enter in the construction of 
different sets of (pure angular momenta) eigenspaces
selected according to the different types of quantum interactions we whish to simulate. 
The fact that physical interactions in many (conservative) quantum systems can be well modelled 
on (combinations of) two--body interactions
\cite{KeBaLi} opens the possibility of calling into play  the powerful 
algebraic--combinatorial setting underlying $SU(2)$ binary coupling and
recoupling theory ({\em cfr.} \cite{BiLo9} and the original references therein).\par 
\noindent Before going into some more details on this realization
of the spin--network graphs, let us point out that
the combinatorial structure encoded into the Racah--Wigner algebra is actually 
shared by other discrete structures.

\begin{figure}[htbp]
\begin{center}
\includegraphics[width=6cm]{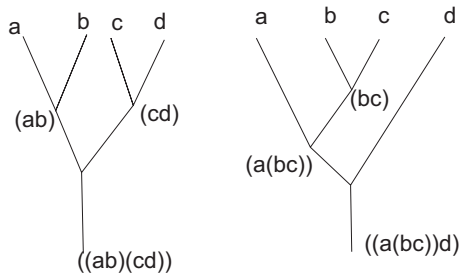}
\end{center}
\caption{Two labelled binary trees on $(n+1)=4$ leaves. Such trees are in one--to--one
correspondence with the vertex set $V$ of the graph $\mathfrak{G}_3(V,E)$}
\label{trees}
\end{figure}

\begin{figure}[htbp]
\begin{center}
\includegraphics[width=6cm]{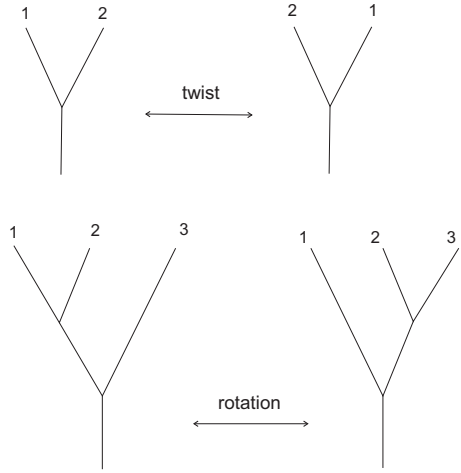}
\end{center}
\caption{A twist corresponds to the interchange of either
two leaves or two subtrees (top). A rotation consists in a change of
the coupling scheme of either three leaves or subtrees (bottom). }
\label{twist-rot}
\end{figure}

A first type of realization is purely graph--theoretical. 
The vertex set $V$ of the graph $\mathfrak{G}_n(V,E)$ 
can be identified with the set of (rooted) binary trees with
$n+1$ labelled leaves where the leaves (terminal nodes) and the internal nodes of the trees 
are labelled by integers and half--integers $\in
\frac{1}{2}\mathbb{N}$, {\em cfr.} Fig. \ref{trees}.
Undirected edges between vertices are drawn whenever a pair of vertices (labelled trees)
are connected by two kinds of topological elementary moves, namely twist and rotation, 
illustrated in Fig. \ref{twist-rot}. 

\begin{figure}[htb]
\begin{center}
\includegraphics[width=7cm]{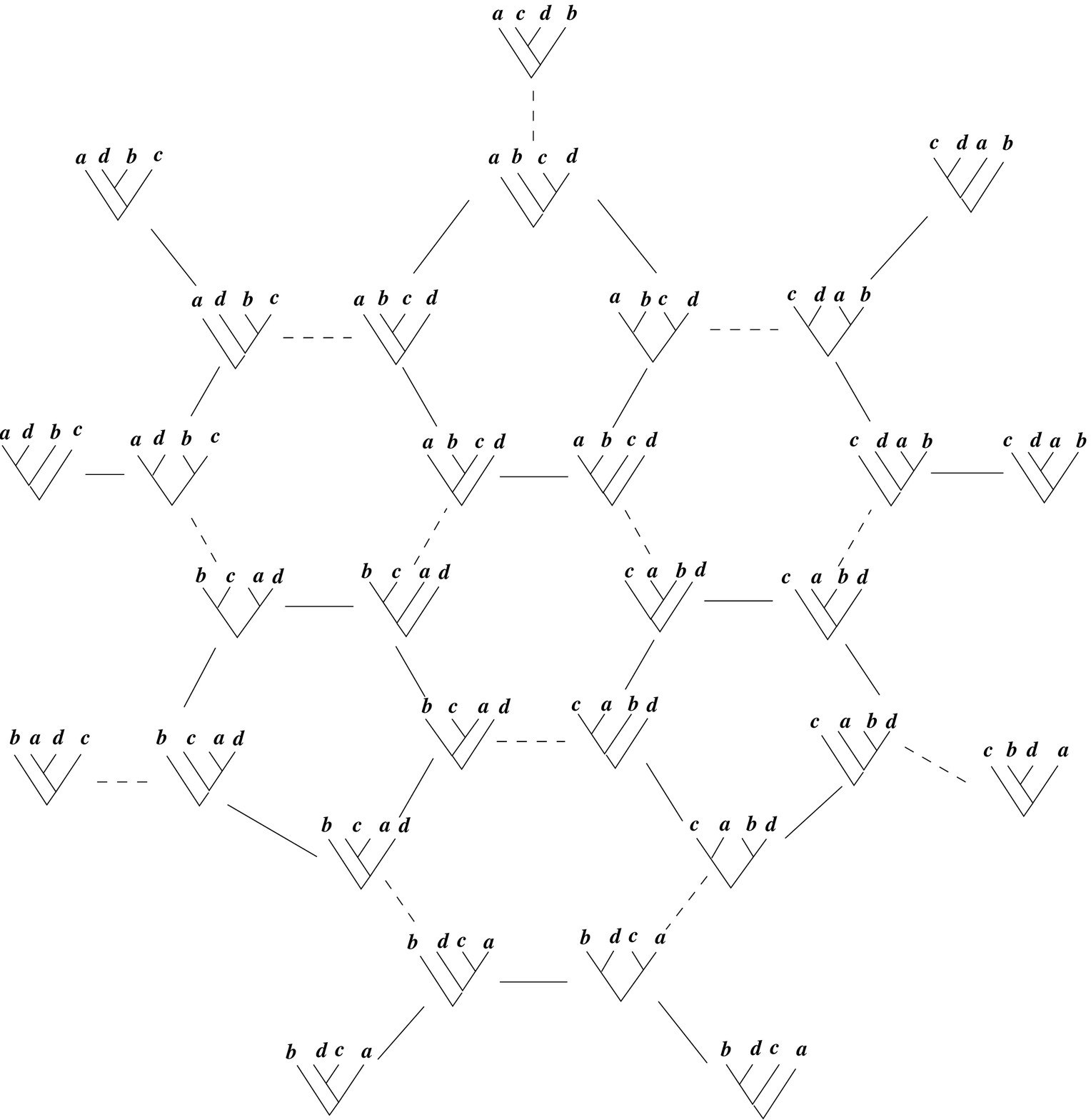}
\end{center}
\caption{A portion of the Twist--Rotation graph $\mathfrak{G}_3(V,E)$ 
where only 30 out of  60 vertices are shown (the picture can be completed by
taking the mirror image of each tree at the antipodal vertex). The remaining 60 vertices 
are arranged into an isomorphic graph obtained 
by swapping one pair of labels, {\em e.g.} $(a,b) \rightarrow (b,a)$. 
Solid edges represent rotations and dashed edges represent twists.}
\label{TRgraph}
\end{figure}

The resulting graph, known as Twist--Rotation graph, is depicted 
for $n+1=4$ in Fig. \ref{TRgraph} and its combinatorial properties 
are analyzed in \cite{AqCo} and in Appendix A of \cite{MaRa2}.

 Another realization of the spin--network is in terms of words 
endowed with pairs of parentheses representing a non--commutative and
non--associative binary operation. In this case the 
vertices of the graph $\mathfrak{G}_n(V,E)$
are associated with  words $w$ made of letters from the alphabet 
$\{$ $\frac{1}{2}\mathbb{N}$ $\cup$ pairs of labelled parentheses $(\cdot \cdot)_a$ $\}$, {\em e.g.}
\begin{equation}\label{parword}
w\,=\,\left( \left( ( j_1 ,j_2)_{k_1} ,j_3  \right)_{k_2},\ldots 
\right)_J \; ; j_i,k_l
\in \tfrac{1}{2}\mathbb{N}\;\;\; \text{with}\; j_1+j_2+\ldots j_{n+1} =J,
\end{equation}
\noindent where $J$ is the label assigned to the root.
Two vertices are connected by an edge if it is possible to switch from one to the other 
either by swapping the elements inside a parenthesis,
$(a,b)_c  \leftrightarrow (b,a)_c$ , 
or by changing the parenthesization structure 
$
\left( (  \cdot\,,  \cdot )_{k_1 } \, , \cdot  \right)_{k_2}\,$ $ \leftrightarrow 
\left(  \cdot \, , ( { \cdot  \, ,
\cdot } )_{h_1 }  \right)_{h_2 } 
$. 

Coming back to the Racah--Wigner setting, the interpretation of the spin--network graph
goes on as follows.
There exists a one--to--one correspondence
$\{v(\mathfrak{b})\}  \longleftrightarrow \{{\cal H}^J_n\,(\mathfrak{b})\}$
between the vertices of $\mathfrak{G}_n(V, E)$ and the computational Hilbert spaces 
of the simulator. 
The label ${\mathfrak{b}}$ has the following meaning: for any given pair $(n, J)$, all binary coupling 
schemes of the $n+1$ angular momenta $\bigl \{ {\bf J}_{\ell} \bigr \}$, identified 
by the quantum numbers $j_1, \dots , j_{n+1}$ (summing up to a total $J$) 
plus $k_1, \dots , k_{n-1}$ (corresponding 
to the $n-1$ intermediate angular momenta $\bigl \{ {\bf K}_{i} \bigr \}$) and by the 
brackets defining the binary couplings, provide the `alphabet' in which quantum 
information is encoded (the rules and constraints of bracketing are instead part of 
the `syntax' of the resulting coding language). 
The Hilbert spaces ${\cal H}^J_n\, 
(\mathfrak{b})$ thus generated are spanned by 
complete orthonormal sets of states with suitable quantum number label set 
such as, {\em e.g.} for $n=3$, $\bigl \{ \bigl ( \bigl (j_1 \bigl (j_2j_3 \bigr )_{k_1}
\bigr )_{k_2}j_4 \bigr )_J$ , $\bigl ( \bigl (j_1j_2 \bigr )_{k'_1} \bigl ( j_3j_4 
\bigr )_{k'_2} \bigr )_J \bigr \}$. 

More precisely, for a given value of $n$, ${\cal H}^J_n(\mathfrak{b})$ is the simultaneous
eigenspace of the squares of $2(n+1)$ Hermitean, mutually commuting angular
momentum operators
${\bf J}_1,\;{\bf J}_2,\;{\bf J}_3,\ldots,{\bf J}_{n+1}\,$ with fixed sum
${\bf J}_1\,+\,{\bf J}_2\,+\,{\bf J}_3\,+\ldots+{\bf J}_{n+1}\;=\;{\bf J}$, of
the intermediate angular momentum operators
${\bf K}_1,\,{\bf K}_2,\,{\bf K}_3,\,\ldots,\,{\bf K}_{n-1}$
and of the operator $J_z$ (the projection of the total angular momentum $\bf{J}$
along the quantization axis). The associated quantum numbers are 
$j_1, j_2,\ldots,j_{n+1};$ $\,J;$ $ k_1,k_2,\ldots,$ $k_{n-1}$ and $M$, where $-J \leq M
\leq +J$ in integer steps.\\ 
If ${\cal H}^{j_1}\otimes$ ${\cal H}^{j_2}\otimes\cdots$ $\otimes 
{\cal H}^{j_{n}}\otimes {\cal H}^{j_{n+1}}$
denotes the factorized Hilbert space, namely the $(n+1)$--fold tensor product 
of the individual eigenspaces of the $({\bf J}_{\ell})^2\,$'s, the operators 
${\bf K}_i$'s represent intermediate angular momenta generated, through Clebsch--Gordan series, 
whenever a pair of ${\bf J}_{\ell}$'s are coupled. As an example, by coupling
sequentially the ${\bf J_{\ell}}$'s according to the scheme
$(\cdots(({\bf J}_1+{\bf J}_2)+{\bf J}_3)+\cdots+{\bf J}_{n+1})$ $={\bf J}$ -- which generates
$({\bf J}_1+{\bf J}_2)={\bf K}_1$,
$({\bf K}_1+{\bf J}_3)={\bf K}_2$, and so on --
we should get a binary bracketing structure of the type
$(\cdots((({\cal H}^{j_1}\otimes{\cal H}^{j_2})_{k_1}$ $\otimes{\cal H}^{j_3})_{k_2}
\otimes$ $\cdots \otimes
{\cal H}^{j_{n+1}})_{k_{n-1}})_J$, where for completeness we add an overall  
bracket labelled by the quantum
number of the total angular momentum $J$. Note that, as far as $j_{\ell}$'s
 quantum numbers are involved, any value belonging to 
 $\{0,1/2,1,3/2,\ldots \}$ is allowed, while the ranges of the $k_i$'s are suitably 
 constrained by Clebsch--Gordan decompositions
 ({\em e.g.} if $({\bf J}_1+{\bf J}_2)={\bf K}_1$ $\Rightarrow$ $|j_1-j_2| \leq$
 $k_1 \leq j_1+j_2$).\\
We denote a binary coupled basis of $(n+1)$ angular
momenta in the $JM$--representation 
and the corresponding Hilbert space as
$$\{\,|\,[j_1,\,j_2,\,j_3,\ldots,j_{n+1}]^{\mathfrak{b}}\, ;k_1^{\mathfrak{b}\,},\,k_2^{\mathfrak{b}\,}
,\ldots,k_{n-1}^{\mathfrak{b}}\, ;\,JM\, \rangle,\;
-J\leq M\leq J \}$$
\begin{equation}\label{bbspace}
=\;{\cal H}^{J}_{\,n}\;(\mathfrak{b})\;\doteq\;\mbox{span}\;\{\;|\,\mathfrak{b}\,;JM\,\rangle_n\,\}\;,
\end{equation}
where  the string inside $[j_1,\,j_2,\,j_3,\ldots,j_{n+1}]^{\mathfrak{b}\,}$ 
 is not necessarily
an ordered one, $\mathfrak{b}$ indicates the current binary bracketing structure and 
the $k_i$'s are uniquely associated with the chain of pairwise couplings selected by $\mathfrak{b}$.\\
For a given value of $J$
each ${\cal H}^J_n (\mathfrak{b})$ has dimension $(2J + 1)$ over 
$\mathbb{C}$, but Hilbert spaces corresponding to 
different bracketing schemes, although isomorphic, are not identical. They actually 
correspond to (partially) different complete sets of physical observables, namely for instance
$\{{\bf J}^2_1,$ $\,{\bf J}^2_2,\,$ ${\bf J}^2_{12},\,{\bf J}^2_3,$ $\,{\bf J}^2,\,J_z\}$ and 
$\{{\bf J}^2_1,\,$ ${\bf J}^2_2,\,{\bf J}^2_3,\,{\bf J}^2_{23},\,{\bf J}^2,\,J_z\}$
respectively (in particular, ${\bf J}^2_{12}$ and ${\bf J}^2_{23}$ cannot be measured 
simultaneously). On the mathematical side this remark reflects the fact that the tensor 
product $\otimes$ is  an associative operation only up to isomorphisms.

For what concerns unitary operations acting on the computational
Hilbert spaces (\ref{bbspace}), we shall consider here unitary transformations 
associated with recoupling 
coefficients ($3nj$ symbols) of $SU(2)$, thought of as $j$--gates in the
present quantum computing context. As  shown in
\cite{BiLo9}, any such coefficient can be 
splitted into `elementary' $j$--gates, namely Racah and phase transforms.
A Racah transform applied to a basis vector is defined formally as
\begin{equation}\label {Cracah1}
{\cal R}\;:| \dots (\,( a\,b)_d \,c)_f \dots;JM \rangle \mapsto
|\dots( a\,(b\,c)_e\,)_f \dots;JM \rangle,
\end{equation} 
where Latin letters $a,b,c,\ldots$ are used here to denote generic, 
both incoming ($j_{\ell}\,$'s
in the previous notation) 
and intermediate ($k_i\,$'s) spin quantum numbers (this operation
corresponds to a rotation in the Twist--Rotation graph, {\em crf.} Fig. \ref{twist-rot},
bottom and Fig. \ref{TRgraph}).
Its explicit expression reads
$$|(a\,(b\,c)_e\,)_f\,;M\rangle$$
\begin{equation}\label{Cracah2} 
=\sum_{d}\,(-1)^{a+b+c+f}\; [(2d+1)
(2e+1)]^{1/2}
\left\{ \begin{array}{ccc}
a & b & d\\
c & f & e
\end{array}\right\}\;|(\,(a\,b)_d \,c)_f \,;M\rangle,
\end{equation}
where there appears the $6j$ symbol of $SU(2)$ and $f$ plays the role
of the total angular momentum quantum number.  Note that, according to the
Wigner--Eckart theorem, the quantum number $M$ (as well as the angular
part of wave functions) is not altered by such 
transformations, and that the same happens with any $3nj$ symbol.
On the other hand, the effect of a phase transform $\Phi$ (a twist operation
on the Twist--Rotation graph, see Fig. \ref{twist-rot}, top and 
Fig. \ref{TRgraph}) amounts to introducing a
suitable phase  whenever two spin labels are swapped
\begin{equation}\label{Cphase}
| \dots ( a\,b)_c \dots;JM \rangle\; = \;(-1)^{a+b-c}
\,|\dots( b\,a)_c \dots;JM \rangle. 
\end{equation}
These unitary operations are combinatorially encoded into 
the edge set $E = \{e\}$ of the graph $\mathfrak{G}_n(V, E)$: $E$ is just
the subset of the Cartesian
product $(V \times V )$ selected by the action of these unitary $j$--gates.

In the framework described above, a computation is 
represented in a natural way by a collection of 
step--by--step transition rules (gates), namely
a family  of `elementary unitary  operations' 
and we assume that it takes
one unit of the intrinsic discrete time variable to perform anyone of them.
Such prescriptions amount to select (families of) `directed paths' in
the spin--network computational space
$\mathfrak{G}_n (V, E)\, \times\, \mathbb{C}^{2J+1}$,
all starting from the same input state and ending 
in an admissible output state. A single path in the given
family can be interpreted as a
(finite--states) quantum automaton calculation, once we select a particular encoding scheme
for the problem we wish to address. 

By a directed path $\cal{P}$ with fixed endpoints
we mean a (time) ordered sequence 
\begin{equation}\label{Cpath}
|\mathbf{v}_{\mbox{in}}\,\rangle_n\equiv
|\mathbf{v}_{0}\,\rangle_n\rightarrow
|\mathbf{v}_{1}\,\rangle_n\rightarrow\cdots\rightarrow
|\mathbf{v}_{s}\,\rangle_n\rightarrow\cdots\rightarrow
|\mathbf{v}_{\mathbb{L}}\,\rangle_n\equiv
|\mathbf{v}_{\mbox{out}}\,\rangle_n\;,
\end{equation}
where we use the shorthand notation $|\mathbf{v}_{s}\rangle_n$ for computational states
(which are vectors expressed in the bases \eqref{bbspace}) and
$s=0,1,2,\ldots ,\mathbb{L}({\cal P})$ is the lexicographical labelling of the states along the  path. 
Finally, $\mathbb{L}({\cal P})$ is the length of the path ${\cal P}$
and $\mathbb{L}({\cal P}) \cdot \tau \doteq T$ is the time required to perform the process in terms of
the discrete time unit $\tau$. 

A computation consists in evaluating the expectation value of the unitary
operator $\mathbf{U}_{{\cal P}}$
associated with the path ${\cal P}$, namely
\begin{equation}\label{snexp1}
\langle \mathbf{v}_{\mbox{out}}\,|\,\mathbf{U}_{\cal{P}}\,|\,
\mathbf{v}_{\mbox{in}}\,\rangle_n.
\end{equation}
By taking advantage of the possibility of decomposing 
$\mathbf{U}_{{\cal P}}$
 uniquely into an ordered sequence of elementary gates, (\ref{snexp1}) becomes
\begin{equation}\label{snexp2}
\langle \mathbf{v}_{\mbox{out}}\,|\,\mathbf{U}_{{\cal P}}\,|\,
\mathbf{v}_{\mbox{in}}\,\rangle_n\;=\;
\lfloor\,
\prod_{s=0}^{\mathbb{L}-1}\,
\langle \mathbf{v}_{s+1}\,|\,{\cal U}_{s,s+1}\,|\,
\mathbf{v}_{s}\,\rangle_n\;\rfloor_{{\cal P}}
\end{equation}
with $\mathbb{L}\equiv \mathbb{L}({\cal P})$ for short. The symbol  
$\lfloor \; \rfloor_{{\cal P}}$ denotes the ordered 
product along the path ${\cal P}$ 
and each elementary operation 
is rewritten as  ${\cal U}_{s, s+1}$ $(s =0,1,2, \ldots \mathbb{L}({\cal P}))$
to stress its `one--step' character. Such expectation values
are  particular instances of the general expression \eqref{autamp} for the quantum amplitude of
a finite--states automaton, once a suitable language
has been encoded into the computational space of the spin--network simulator.

\section{$q$--braided computational space}

\begin{figure}[htb]
\begin{center}
\includegraphics[width=7cm]{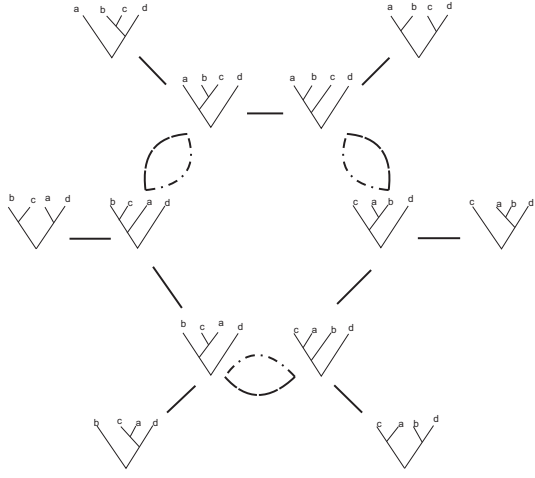}
\end{center}
\caption{A portion of the $q$--braided Twist Rotation graph $(\mathfrak{G}_3 (V, E) \times
\mathbb{Z}_2)_q$: with respect to the unbraided situation, each twist
has been splitted.}
\label{brTRgraph}
\end{figure}

As we shall see in the following section, the basic ingredient for addressing 
link invariants arising in the context of Chern--Simons field theory
is the `tensor structure' naturally associated with the representation ring
of the Lie algebra of a simple compact group which plays the role of
the gauge group of the theory.
In the case of $SU(2)$ this structure is provided by (tensor products of) Hilbert spaces
supporting irreducible representations together with unitary morphisms between them:
these are exactly the objects collected into the Racah--Wigner algebra 
discussed in section 3. However,
when dealing with (planar diagrams of) links we shall also have to specify the 
eigenvalues of the braiding matrix to be associated with the crossings of the links
and this extension can be achieved by `braiding' the Racah--Wigner tensor category. In the present
context, it is natural to take advantage of quantum group techniques in order to `split' any
phase transform (\ref{Cphase}) by assigning different weights --depending
on a deformation parameter $q$ to be defined below-- to right and
left handed twists. From the combinatorial viewpoint, this generalization corresponds to
replace the spin network computational space 
$\mathfrak{G}_n (V, E)\, \times\, \mathbb{C}^{2J+1}$ 
with its $q$--braided counterpart (see Fig. \ref{brTRgraph})
\begin{equation}\label{qgraph}
\left( (\mathfrak{G}_n (V, E)\, \times\, \mathbb{C}^{2J+1})\;\times \,
\mathbb{Z}_2\right)_q, 
\end{equation}
where the (classical) $6j$ symbol in any Racah trasform (\ref{Cracah2}) 
will become $q$--deformed.

The tensor category we are going to introduce is associated with the quantum group
$U_q(su(2))$ ($q =$ root of unity),  namely the universal enveloping algebra of $SU(2)$ 
endowed with additional structures which make it a 
quasitriangular quasi--Hopf--*algebra (see {\em e.g.} 
\cite{Ka} and other references therein).
$U_q \left( {su\left( 2 \right)} \right)$ 
is an associative algebra generated by elements 
$J_+$, $J_-$ and $J_z$ which satisfy the commutation relations
\begin{equation}\label{qcomm}
\left[ {J_z ,J_ \pm  } \right] =  \pm J_ \pm  ;
\left[ {J_ +  ,J_ -  } \right] = \left[ {2J_z } \right]_q , \hfill \\
\end{equation} 
where the $q$--integer $[\mathbf{n}]_q$ is defined as 
$[\mathbf{n}]_q  \equiv (q^{\mathbf{n}/2}-q^{-\mathbf{n}/2})/
(q^{1/2}-q^{-1/2})$.
$U_q(su(2))$ is a deformation of the universal 
enveloping algebra of the Lie algebra $su(2)$ since  in the limit $q\rightarrow 1$ 
the above relations reduce to the commutation relations for the $su(2)$ generators. \\
A Hopf algebra--structure can be  
introduced by defining  the coproduct homomorphism
$$
\Delta :U_q (su(2) )\to U_q (su(2) )\otimes U_q (su(2) ),
$$
acting on $J_+$, $J_-$ and $J_z$ according to
$$
\Delta \left( {J_\pm } \right)=J_\pm \otimes 
q^{\frac{J}{2}}+q^{-\frac{J}{2}}\otimes J_\pm;
$$
$$ 
\mbox{ }\Delta (J_z)=J_z\otimes 
1+1\otimes J_z.
$$
The tensor algebra associated with $U_q(su(2))$ can be worked out in practice as in the case of $su(2)$,
so that we have Hilbert spaces supporting irreducible representations, $q$--Clebsch--Gordan coefficients,
$q$--Racah coefficients and so on. The crucial difference consists in the fact that the irreps label set
acquires a cut--off, namely each label must be chosen 
in the set $\{0,1/2,1,3/2,\ldots,$ $k-2\}$, where the integer $k$ is related to the deformation parameter
$q$ by $q$ $=\exp (-2i \pi /k)$.
\noindent Denoting by $\mathcal{H}_q^{j_1}$ and $\mathcal{H}_q^{j_2}$ the Hilbert spaces supporting two irreps
$j_1,j_2$,
their (truncated) tensor product can be decomposed according to the Clebsch--Gordan series
\begin{equation}\label{Qseries}
\mathcal{H}_q^{j_1 } \otimes \mathcal{H}_q^{j_2 } \;=\;
\bigoplus_{j=\left| {j_1  - j_2 } \right|}^{\min \left\{ j_1  + j_2 ,k - j_1  - j_2  
\right\}} \;\,\mathcal{H}_{\,q}^{\,j}\; .
\end{equation} 
As happens in the classical case,  
the two bases associated with the eigenspaces involved in the  tensor product \eqref{Qseries} 
can be connected by means of Clebsch--Gordan coefficients according to
\begin{equation}\label{Qclebsch}
\left| {j\,m} \right\rangle_q  = \sum\limits_{m_1 ,m_2 } (j_1j_2m_1m_2\,|\,jm)_q\,\left| 
j_1 \, m_1  \right\rangle_q \left| j_2 \,m_2  \right\rangle_q ,
\end{equation}
where $-j_1 \leq m_1 \leq j_1$, $-j_2 \leq m_2 \leq j_2$, $m=m_1+m_2$.
Quantum CG coefficients $(\;)_q$ can be suitable normalized and satisfy orthogonality relations 
\cite{KiRe}.\\
The quantum Racah transformation comes out when we  consider different 
binary couplings in the tensor product 
$\mathcal{H}_q^{j_1 } \otimes \mathcal{H}_q^{j_2 } \otimes \mathcal{H}_q^{j_3 }$ of three 
irreducible representations, as done in the classical case 
({\em cfr.} \eqref{Cracah1} and \eqref{Cracah2}). For instance
$$
|( j_1 j_2)_{j_{12}}  j_3 ;jm \rangle_q 
$$
\begin{equation}\label{Qracah}
=\sum\limits_{j_{23} } W_q \,( j_1 j_2 j j_3 ;\,j_{12} j_{23}) 
\left(\; [ 2j_{12}  + 1 ]_q [ 2j_{23}  + 1]_q \;\right)^{-1/2}\,
|j_1 ( j_2 j_3 )_{j_{23} } ;jm\rangle_q , 
\end{equation}
where there appear the $q$--dimensions of the irreps involved.
The components of $W_q$ are the Racah coefficients of the algebra $U_q(su_2)$
and these symbols satisfy orthogonality relations, symmetry properties and
identities which look like suitable $q$--deformations of the corresponding classical ones
(and reduce to them in the limit $q \rightarrow 1$) \cite{KiRe}.
The quantum Racah coefficient and the $q$--counterpart of the Wigner $6j$ symbol 
differ as usual by a phase factor, namely 
$$W_q (j_1j_2 j j_3;j_{j_{12}}j_{j_{23}})\,\doteq\
(-1)^{j_1  + j_2  + j_3  + j} \;
\begin{Bmatrix}
j_1 & j_2 & j_{12}\\
j_3 & j & j_{23}
\end{Bmatrix}_q.
$$
Finally, we introduce the (differently normalized) symbol 
\begin{equation}\label{norQracah}
\left( \begin{array}{cc}
j_1&j_2\\
j_3&j
\end{array} \right. \bigg\vert \left. \begin{array}{c}
j_{12}\\
j_{23}
\end{array} \right)_q
\;\doteq\;
 \frac{W_q (j_1j_2 j j_3;j_{j_{12}}j_{j_{23}})}
{{\sqrt {\left[ {2j_{12}  + 1} \right]_q \left[ {2j_{23}  + 1} \right]_q } }},
\end{equation}
which, on the one hand, enhances the matrix character of the quantum Racah 
transform \eqref{Qracah}
and, on the other, is particularly suitable to be generalized
to deal with more than three incoming spin labels.

\section{Quantum invariants of links and
 unitary\\
 representations of the braid group}
 
Link invariants are functions on links (collections of knots, namely closed circles 
in $3$--space) which depend only on the isotopy class of the link.
An (ambient) isotopy can be thought of  as a continuous transformation performed on the link
embedded in $\mathbb{R}^3$ which deformes at will the shape of the link without
cuttings.
 Let us point out preliminarly that the link  
invariants of polynomial--type we are going to address here
are `universal' in the sense that historically distinct approaches (R--matrix representations
obtained with the quantum group method, monodromy representations of the braid group
in $2D$ conformal field theories, the quasi tensor category approach by Drinfeld and the
$3D$  quantum Chern--Simons theory, see {\em e.g.} \cite{Gua,LKau} 
for reviews) are indeed different
aspects of the same underlying algebraic structure. 
We shall focus in particular on the Chern--Simons setting \cite{Wit} since, on the one hand,
it embraces the universal structure of  (unitary) braid group representations shared by all
the models quoted above and, on the other, can be naturally encoded into the (braided)
spin--network computational scheme. 
The (colored) link polynomials arising 
from $SU(2)$ quantum CS theory can be referred to as 
`extended' Jones polynomials, since the Jones polynomial 
\cite{VJo} is recovered 
by selecting the fundamental ($j=\frac{1}{2}$) representation of $SU(2)$ 
on each of the link components
(or on each strand of the associated braid). Moreover, the topological quantum
field approach is inherently related to low dimensional geometry
since, for instance, suitable combinations of these invariants can be 
interpreted as topological invariants  
of hyperbolic $3$--manifolds, obtained by surgery along framed links in the $3$-sphere 
\cite{ReTu,KiMe}. \\
Let us point out that the definitions
of link polynomials from Hecke (or Temperley--Lieb) algebra realizations
of the braid group --exploited in \cite{AhJoLa} in the quantum computational context--
can be derived quite easily  in the framework we are adopting here, since
it can be shown that the associated invariants do satisfy the linear skein relations
which characterize such realizations \cite{Gua}.

Before addressing a full fledged approach to $3D$ Chern--Simons theory, we pause a little bit
digressing on $R$--matrix representations of the braid group arising from quantum groups.
The associated invariants of knots and links
are commonly refereed to as `quantum' invariants, since they are quantities depending
on the deformation parameter $q$ of the `quantum group' under consideration.\par

\subsection{The quantum group approach}

\noindent Let $\mathfrak{g}$ be a (semi)simple Lie algebra, $U_q(\mathfrak{g})$ 
its universal enveloping algebra 
and $V$ a finite dimensional (complex) vector space in the associated tensor algebra
(the prototype is of course the unitary tensor algebra of $U_q (su(2))$ 
described in details in section 4). 
The representation 
theory of any such quantum group is naturally endowed with 
an invertible linear operator, the so--called $R$--matrix
\begin{equation}\label{Rmatrix} 
R:V\otimes V\rightarrow V\otimes V, 
\end{equation} 
which satisfies the quantum Yang--Baxter equation
\begin{equation}\label{QYB}
(R\otimes I)(I\otimes R)(R\otimes I)=(I\otimes R)(R\otimes I)(I\otimes R),
\end{equation}
where both sides of the above expression are to be understood as linear transformations
$
V\otimes V\otimes V\rightarrow V\otimes V\otimes V
$.

\begin{figure}[htbp]
\begin{center}
\includegraphics[width=5cm]{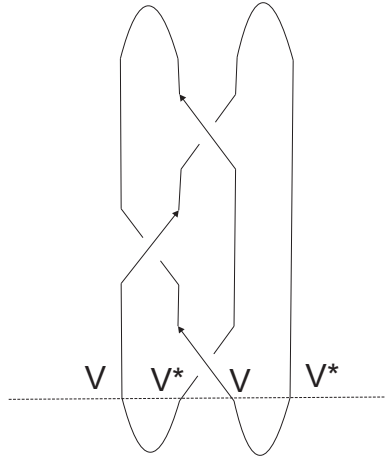}
\end{center}
\caption{The oriented trefoil knot cut by an horizontal line. We associate with
the ordered set of the intersection points (from left to right) the 
tensor product $V\otimes V^*\otimes V\otimes V^*$, where each factor is
chosen  in order to comply with the diagram orientation.}
\label{trefoil}
\end{figure}

The general procedure for constructing quantum invariants of oriented 
knots (or links) presented as closure (or platting) of braids can be outlined as follows. 
Consider an oriented  knot diagram (namely the projection of a knot
with orientation onto
a fixed plane) and insert an  horizontal line as depicted in Fig.
\ref{trefoil}. 

To each intersection point between the line and the diagram we 
assign either the representation space $V$ or its dual $V^*$, depending 
on whether the portion of the knot nearby   
the intersection is oriented upwards or downwards. 
The whole configuration of such points on the line turns out to be associated with the  
tensor product of the individual vector spaces (orderered from left to right).
The connection with  braid groups comes out when we consider two
parallel horizontal lines intersecting the knot diagram.
More precisely, the portion of the knot diagram between a pair of 
horizontal lines represents the geometric realization of a braid $b$, 
which in turn
is an element of the Artin braid group $\mathbf{B}_n$, for some suitable $n$. 
$\mathbf{B}_n$ has $n$ generators, denoted by $\{\sigma_1,\sigma_2,\ldots,
\sigma_{n-1}\}$ plus the identity $e$, which satisfy the relations
$$
\sigma_i\,\sigma_j\,= \,\sigma_j\,\sigma_i \;\;\;\;\mbox{if}\;\,\,|i-j| > 1 \,\;\;(i,j=1,2,\ldots , n-1)
$$
\begin{equation}\label {brgroup}
\sigma_i\,\sigma_{i+1}\,\sigma_i\,= \,\sigma_{i+1}\,\sigma_{i}\,\sigma_{i+1}
 \;\;\;(\,i=1,2,\ldots ,n-2).
\end{equation}
An element of the braid group is a word in the standard generators of $\mathbf{B}_n$,
{\em e.g.} $b= \sigma_3^{-1}\sigma_2$ $\sigma_3^{-1}\sigma_2 \, 
\sigma_1^{3}$ $\sigma_2^{-1}\sigma_1\sigma_2^{-2}$ $\in \mathbf{B}_4$; the length $|b|$
of the word $b$ is the number of its letters.
The group acts naturally on topological sets of $n$ disjoint strands --
ordered from left to right --
in the sense that each generator $\sigma_i$ corresponds to the over--crossing of the
$i$th strand on the $(i+1)$--th,
and $\sigma_i^{-1}$ represents the inverse operation (under--crossing)
according to $\sigma_i\,\sigma_i^{-1}$ $=\sigma_i^{-1}\sigma_i =e$.

\begin{figure}[htbp]
\begin{center}
\includegraphics[width=6cm]{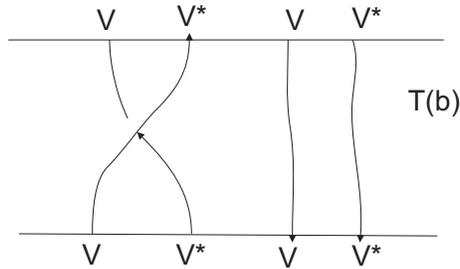}
\end{center}
\caption{The action of the braid group element $b$ is represented as a map $T(b)$ between 
the vector spaces living on the bottom and top lines.}
\label{braidmap}
\end{figure}

On the other hand, when we represent $\mathbf{B}_n$ in the tensor algebra of
$U_q(\mathfrak{g})$, the action of a braid $b$ is naturally associated with 
a linear operator 
$T(b)$ connecting the vector spaces introduced above, see Fig.
\ref{braidmap}. 

\noindent Since $T$ is a linear representation, we can simply specify its action on the standard
generators $\{\sigma_1,\sigma_2,\ldots,$
$\sigma_{n-1}\}$ to get $\{T(\sigma_1),T(\sigma_2),$ $\ldots,
T(\sigma_{n-1})\}$, and extend this action to $T(b)$ by linearity.
The $R$--matrix, namely the linear operator introduced in \eqref{Rmatrix},
is to be intended as the set of (elementary) crossing operators in some
given representation $T$, constrained by the quantum Yang--Baxter relation
\eqref{QYB}.

Knot theory is closely related to (representations of) braid groups owing to 
Alexander's theorem \cite{Alex}, which states that every knot (or link ) $L$ in the $3$--sphere $S^3$
$=\mathbb{R}^3 \cup \infty$ can be
presented (not uniquely) as a closed braid for some suitable $n$
(to get a knot from the open braid of Fig. \ref{braidmap} we have to
connect with arcs the lower and upper endpoints of each strand). 
We might also consider the plat presentation of a knot (characterized by
the fact that the braid involved must possess an even number of strands), which is
exactly the type of presentation depicted in Fig. \ref{trefoil} for the trefoil knot
(see also Fig. \ref{borromeanplat} and Fig. \ref{trefoilEx} in the Appendix).
 Anyway, we can generate invariants 
of knots (links) for both types of presentations by taking some 
'trace' of the operator $T(b)$, where $b$
is the braid associated with the given knot or link. The fact that the
resulting quantities must depend only on the isotopy type of the knot 
can be suitably translated into
the braid group--setting by resorting to the notion of invariance under Markov
moves, and thus we should actually  speak of `Markov traces'
({\em cfr.} \cite{BiBr,Bir} for reviews on knot 
theory and braid group). 

Summing up, the quantum group approach provides a purely algebraic construction of link invariants
as (Markov) traces of representation matrices of the braid group in the
tensor algebra of $U_q(\mathfrak{g})$. Such invariants are polynomials
in the deformation parameter $q$ and its inverse $1/q$. In the case 
of $U_q (su(2))$ ($q$ a root of unity), the associated $q$--braided Racah--Wigner 
algebra (discussed in section 4) is naturally endowed with Hilbert spaces and 
unitary operators, namely the ideal arena
to address (quantum) computational problems concerning both link polynomials and
braid group.

\subsection{The Chern--Simons field theory approach} 
A topological quantum field theory (TQFT) is a particular type of gauge theory,
namely a theory quantized through the
(Euclidean) path integral prescription starting from a classical Yang--Mills action 
defined on a suitable $D$--dimensional space(time).
TQFT are characterized by observables 
(correlation functions) which depend only on the
global features of the space on which these theories live, namely they
are independent of any metric which may be used to define the underlying classical theory.
The geometrical generating functionals
and correlation functions of such theories are computable by standard techniques in quantum field
theory and provide novel representations of certain global invariants (for $D$-manifolds and/or 
for particular submanifols embedded in the ambient space) which are of prime interest.
 Let us recall in brief the basic axioms for a unitary TQFT in $D=3$ before
going through the case which is  of interest here, namely $SU(2)$ Chern--Simons theory
\cite{Ati}.

Denote by $\Sigma_1$ and $\Sigma_2$ a pair of 2--dimensional manifolds and by
$\mathcal{M}^3$ a $3$--dimensional manifold with boundary $\partial \mathcal{M}^3$
$=\Sigma_1 \cup \Sigma_2$ (all manifolds here are compact, smooth and oriented). 
A unitary $3$--dimensional quantum field theory corresponds to the assignment of \\
{\bf i)} finite dimensional Hilbert spaces (endowed with non--degenerate
bilinear forms) $\mathcal{H}_{\Sigma_1}$ and $\mathcal{H}_{\Sigma_2}$
to $\Sigma_1$ and $\Sigma_2$, respectively;\\
{\bf ii)} a map (technically, a functor) connecting such Hilbert spaces
\begin{equation} \label{TopFunct}
\mathcal{H}_{\Sigma_1}\;
\xrightarrow{\mathbf{Z}\,[\mathcal{M}^3\,]}\;
\mathcal{H}_{\Sigma_2}
\end{equation}
\noindent where $\mathcal{M}^3$ is a manifold which interpolates between $\Sigma_1$
(incoming boundary) and  $\Sigma_2$ (outgoing boundary). 
Without entering into details concerning a few more axioms
(diffeomorphism invariance, factorization {\em etc.}) we just recall that unitarity
implies that\\
{\bf iii)} if $\bar{\Sigma}$ denotes the surface $\Sigma$ with the opposite orientation, then
$\mathcal{H}_{\bar{\Sigma}}=$ 
$\mathcal{H}^{*}_{\Sigma}$, where $*$ stands for complex conjugation;\\ 
{\bf iv)} the mappings \eqref{TopFunct} are unitary and 
$\mathbf{Z}[\bar{\mathcal{M}}^3]=$ 
$\mathbf{Z}^{*}[\mathcal{M}^3]$, where  $\bar{\mathcal{M}}^3$ denote the manifold
with the opposite orientation with respect to $\mathcal{M}^3$. 

The classical  $SU(2)$ Chern--Simons action for the sphere $S^3$ 
(which is the simplest compact, oriented $3$--manifold without boundary) is given by
\begin{equation}\label{CSaction}
k\,S(A)=\frac{k}{4\pi}\int_{S^3}tr(AdA+\frac{2}{3}A \wedge A \wedge A)
\end{equation}
where $A$ is the connection 1--form with value in the Lie algebra  
$su(2)$ of the gauge group, $k$ is the coupling constant, $d$ is the exterior
differential, $\wedge$ is the wedge product of
differential forms and the trace is taken over Lie algebra indices.
The partition function of the quantum theory
corresponds to the map \eqref{TopFunct} restricted to the case of
empty boundaries and  is obtained as a `path integral', namely
by integrating the exponential
of $i$ times the classical action \eqref{CSaction} over the space of gauge--invariant 
flat $SU(2)$ connections (the field variables) according to the formal expression 
\begin{equation} \label{CSfunct}
\mathbf{Z}_{\,CS}\,[S^3; k]\;=\;
\int [DA]\,\exp \left\{\frac{i\, k}{4 \pi}\,
S_{CS}\,(A)\,\right\}
\end{equation}
where the coupling constant $k$ is constrained to be 
a positive integer by the gauge--invariant quantization procedure
and is related to the deformation parameter $q$ (see below).
The generating functional \eqref{CSfunct}, written for a 
generic compact oriented $3$--manifold 
$\mathcal{M}^3$ with $\partial \mathcal{M}^3 = \emptyset$, is a global invariant,
namely depends only on the topological type \cite{Wit}.

The extension of \eqref{CSfunct} to the case of a manifold with boundaries,  
$\partial \mathcal{M}^3 \neq \emptyset$,
requires modifications
of the classical action \eqref{CSaction} by suitable 
Wess--Zumino--type terms to be associated with each boundary component \cite{Car}. However, 
we do not need here the explicit expression of such boundary action 
since what we are interested
in are expectation values of observables in the quantized  field theory
which will just require the knowledge of (vectors belonging to) the boundary Hilbert spaces, 
{\em cfr.} {\bf i)} above. 
In particular, it turns out that the gauge--invariant 
observables in the quantum CS theory are expectation values of Wilson line operators 
associated with oriented knots (links) embedded
in the $3$--manifold (commonly referred to as Wilson loop operators).
Knots and link are `colored' with irreps of the gauge group $SU(2)$, restricted 
to values ranging over $\{0,1/2,1,3/2,\ldots, k-2\}$, where the integer $k$ is 
related to the deformation parameter $q$ by $q= \exp(-2i\pi/k)$
(see section 4 for details on the $U_q (su(2))$ representation algebra).

The Wilson loop operator associated with a knot
$K$ carrying a spin--$j$ irreducible representation is defined as (the trace of)
the holonomy of the connection 1--form $A$ evaluated along the closed loop
$K$ $\subset S^3$, namely
\begin{equation}\label{WilK}
\mathbf{W}_j\,[K]=tr_j\,Pexp\oint_KA,
\end{equation}
where $P$ is the path ordering.
For a link $L$ made of a collection of $s$ knots $\{K_l\}$, each labelled by an
irrep, the expression of the composite Wilson operator reads
\begin{equation}\label{WilL}
\mathbf{W}_{j_1j_2\ldots j_s}\,[L]\,=\,\prod_{l=1}^s\;\mathbf{W}_{j_l}\,[K_l].
\end{equation}
In the framework of the path integral quantization procedure, 
expectation values of observables are defined as functional
averaging  weighed with the exponential of the classical action.
In particular, the functional average of the Wilson operator
\eqref{WilL} is
\begin{equation}\label{Wilexpect}
\mathcal{E}_{j_1...j_s}\,[L]\,=\,\frac{
\int [DA]\;\mathbf{W}_{j_1 \ldots j_s}\,[L]\,e^{\,\frac{ik}{ 4\pi}
\,S_{CS}\,(A)}} {\int [DA]\;e^{\,\frac{ik}{ 4\pi}
S_{CS}\,(A)}},
\end{equation}
where $S_{CS}\,(A)$ is the CS action for the $3$--sphere given in \eqref{CSaction}
and the generating functional in the denominator will be normalized
to 1 in what follows.
It can be shown that this expectation value, which essentially\footnote{These 
polynomials are actually invariants of `regular' isotopy, which represents a restricted
form of `ambient' isotopy defined at the beginning of this section. The connection
between $\mathcal{E}_{j_1...j_s}\,[L]$ and the genuine colored Jones polynomial is given by
$J_{j_1...j_s}(L,q^{1/2})= $ $\{q^{-3\mathit{w}(L)/4}/(q^{1/2}-q^{-1/2})\}$
$\mathcal{E}_{j_1...j_s}\,[L]$, once suitable normalizations for the unknots 
have been chosen. Here $\mathit{w}(L)$ is the writhe associated with the planar 
diagram $D(L)$ of the link $L$, defined as $\mathit{w}(L)=\sum_p \varepsilon (p)$.
The summation runs over the self crossing points of $D(L)$ and $\varepsilon (p)=\pm 1$ 
according to simple combinatorial rules (see {\em e.g.} \cite{LKau}).
The writhe is easily recognized from the link diagram by simple counting arguments, 
so that computational problems involving  both link invariants 
belong to the same complexity class.}
coincides with the extended (colored) Jones polynomial
\cite{ReTu,KiMe}, 
depends only on the isotopy type of the oriented 
link $L$ and on the set of irreps $\{j_1,...,j_s\}$
(note also that $\mathcal{E} [L] =\mathcal{E} [\bar{L}]$, where 
$\bar{L}$ is obtained from $L$ by reversing the orientation).

\begin{figure}[htbp]
\begin{center}
\includegraphics[width=6cm]{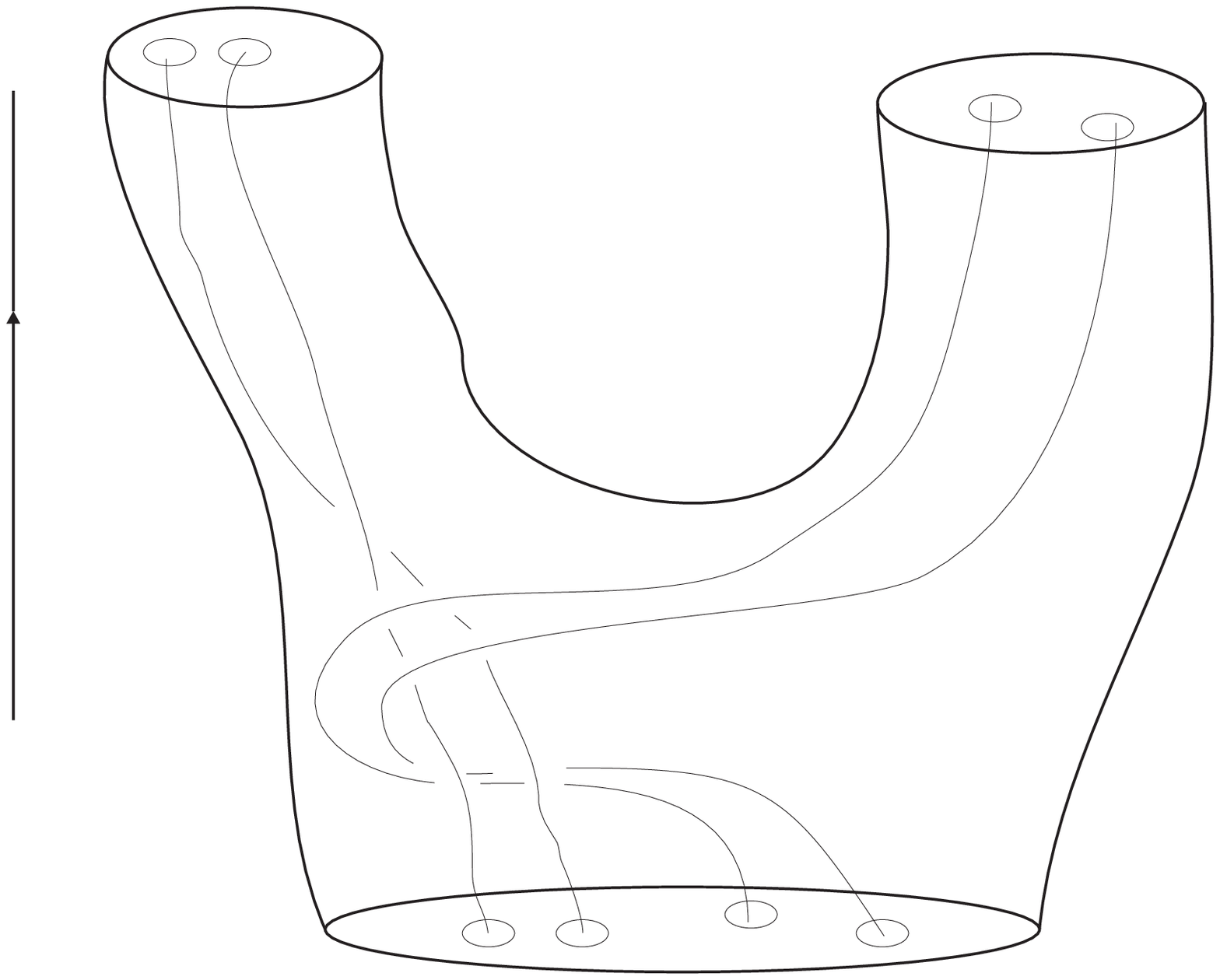}
\end{center}
\caption{A portion of an oriented  $3$--manifold with one incoming
boundary and two outgoing boundaries. Lines belong to some knot (or link)
embedded in the manifold and intersect the $2D$ boundaries in some
points (punctures).}
\label{pant}
\end{figure}

The explicit evaluation of \eqref{Wilexpect} can be carried out 
in several ways, by resorting to either
field--theoretic methods, quantum group  approaches 
(outlined above) or through combinatorial state sum functionals. 
For future convenience we just sketch here the approach which 
relies on the extension of CS quantum theory --endowed with a Wess--Zumino
conformal field theory on its boundary-- to the case in which the boundary
components are intersected by knots or links, namely become
$2$--manifolds with punctures (note that this setting  is 
closely related to the topological quantum computation approach
\cite{FrKiLaWa}). The basic geometric ingredients can be easily
visualized  as in Fig.
\ref{pant},
where a portion of a $3$--dimensional manifold $\mathcal{M}^3$ 
(technically, a handlebody decomposition) is shown, together with an incoming boundary
$\Sigma_1$ and an outgoing boundary $\Sigma_2$ made of two disjoint
components, $\Sigma^{'}_2$ and $\Sigma^{''}_2$. A portion of some
knot (link) embedded in the ambient $3$--manifold is also depicted,
and its intersections with the boundaries are `punctures' which inherit
the irreps labels from the associated (Wilson) lines.   

According to the axioms of TQFT, we may associate with each boundary a (finite--dimensional)
Hilbert space, that is $\mathcal{H}_{\Sigma_1}$ for the incoming boundary and
$\mathcal{H}_{\Sigma_2}\doteq \mathcal{H}_{\Sigma^{'}_2}\otimes \mathcal{H}_{\Sigma^{''}_2}$
(here, for simplicity, we do not explicitate the labels of puncures).
The Chern--Simons unitary functional (see axiom {\bf ii)}) is
a state in the tensor product of these Hilbert spaces or, more precisely,
$$
\mathbf{Z}_{CS}\,[\mathcal{M}^3\,;k]\;:\;\mathcal{H}_{\Sigma_1}\;\rightarrow
\mathcal{H}_{\Sigma_2}
$$
\begin{equation}\label{CStensor}
\Rightarrow \;\; \mathbf{Z}_{CS}\,[\mathcal{M}^3\,;k]\;\;\in\;\mathcal{H}_{\Sigma_1}\,\otimes
\mathcal{H}^{*}_{\Sigma_2},
\end{equation}
where in the last row we have used also axiom {\bf iii)} since the incoming and outgoing boundaries 
must be endowed with opposite orientations. Moreover,
such type of expression is compatible with the quantum group approach outlined in section 5.1
since the the Chern--Simons mapping
in \eqref{CStensor}, when restricted to punctures, induces automatically
(unitary) representations of the braid group in the tensor algebra of
$U_q(su(2))$.\\
Finally, it can be shown \cite{Ati} that the conformal blocks 
of the $SU(2)_{\ell}$ Wess--Zumino field theory living  on the boundaries with 
punctures actually provide the basis vectors for the Hilbert spaces introduced above 
(the level $\ell$ of the WZ model is related to the deformation
parameter $q$ according to $q = \exp \{- 2\pi i/(\ell + 2)\}$, and in turn $\ell$
is related to the coupling constant $k (\geq 3)$ of the CS theory in the bulk by $\ell=k-2$).
In the following section we shall carry on the
explicit construction of such bases, which will allow us to recast
the expectation value of the composite Wilson operator \eqref{Wilexpect}
into a form suitable to be handled for computational purposes.

\section{Automaton calculation of extended\\
 Jones polynomials}

As anticipated at the end of the previous section,  
we start with the construction of  the basis vectors
which will enter into the explicit expression of the expectation
value of the composite Wilson operator \eqref{Wilexpect}
emerging from quantized $3D$ Chern--Simons theory.
We use here the general setting given in \cite{Kaul} since  
it can be  easily adapted  to the $q$--braided
spin--network scheme of section 4.\\
Consider an oriented link $L$ embedded in the $3$--sphere, $S^3$ 
$= \mathbb{R}^3 \cup \infty$,
endowed with a plat representation, namely presented as
the closure of an oriented braid with $2N$ strands 
({\em cfr.} Figg. \ref{borromeanplat},
\ref{trefoil} and \ref{trefoilEx}).
If we remove two open three--balls from $S^3$ we get two boundaries, $\Sigma_1$
and $\Sigma_2$, both topologically equivalent to $S^2$,
but with opposite orientations, $(S^3;\Sigma_1,\Sigma_2)$ $\equiv$ 
$(S^3;S^2,\bar{S}^2)$ (recall from section 5.2 that an $SU(2)_{\ell}$ Wess--Zumino conformal
field theory is naturally associated with the oriented boundary surfaces). 
We can accomodate in such an ambient manifold,
$2N$ 
`unbraided' Wilson lines carrying irreps $j_1,j_2,\ldots,j_{2N}$,
starting from the incoming (lower) boundary and ending into the outgoing (upper) one
(punctures inherit the labellings $j_i \in \{0,1/2,1,3/2,\ldots \ell\}$ from the 
strands of the braid).   Denote this `identity'  colored oriented braid as
\begin{equation}\label{idbraid}
\nu_{\,I}\;\left( \begin{array}{cccc}
   \widehat{j}_1^*  & \widehat{j}_2^*  & \ldots & \widehat{j}_{2N}^*   \\
   \widehat{j}_1  & \widehat{j}_2  & \ldots & \widehat{j}_{2N}   
\end{array} \right),
\end{equation}
where $\widehat{j}_i \equiv (j_i, \epsilon_i)$ $i=1,2,\ldots 2N$ represents the spin $j_i$ together with
an orientation $\epsilon_i\,=\pm 1$ for a strand going into or away from the boundary,
while stars over the symbols represent here the opposite choice of the
orientation.  \\
In order to generate an arbitrary (oriented) braid $\nu_B$ out of the identity braid
$\nu_I$ we have to apply a braiding operator, denoted by the  symbol
$B$ and written in terms of generators $B_1,B_2,...,B_{2N-1}$ to be defined below,
starting from the lower boundary. With such prescription we shall get the braid
\begin{equation}\label{genbraid}
\nu _{\,B}\; \left( {\begin{array}{ccccc}
   {\widehat{j}_1 } & {\widehat{j}_1^* } & {\ldots} & {\widehat{j}_N } & {\widehat{j}_N^* }  \\
   {\widehat{l}_1 } & {\widehat{l}_1^* } & {\ldots} & {\widehat{l}_N } & {\widehat{l}_N^* }  \\
 \end{array} } \right) , 
\end{equation}
where the labels have been ordered according to the requirement of having a plat
presentation for the associated oriented link. 

\begin{figure}[htbp]
\begin{center}
\includegraphics[width=6cm]{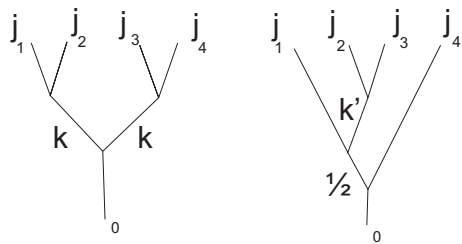}
\end{center}
\caption{Combinatorial  realization of the two basis sets in the case
$N=4$ as labelled  binary trees. They are connected by a duality matrix.}
\label{RacahOnTrees}
\end{figure}

Our goal will consist in recasting the expectation value of the composite Wilson operator,
written in  functional terms in \eqref{Wilexpect}, into an expression which
contains a quantity of the type 
\begin{equation}\label{first}
\langle\, \phi \,\lvert\,B\, \left( {\begin{array}{ccccc}
   {\widehat{j}_1 } & {\widehat{j}_1^* } & {\ldots} & {\widehat{j}_N } & {\widehat{j}_N^* }  \\
   {\widehat{l}_1 } & {\widehat{l}_1^* } & {\ldots} & {\widehat{l}_N } & {\widehat{l}_N^* }  \\
 \end{array} } \right)\,\rvert
\,\tilde{\phi} \,\rangle,
\end{equation}
where $B(\;:::\;)$ is the operator associated with the oriented braid \eqref{genbraid}.
 The shorthand notations $|\phi>$ and $|\tilde{\phi}>$ represent 
correlators of
$2N$ primary fields in the $SU(2)_{\ell}$ WZ theory, to be interpreted here as
states belonging to the boundary Hilbert spaces associated, respectively, with the incoming
and outgoing Hilbert spaces of the underlying CS theory ({\em cfr.} the axioms for TQFT 
in section 5.2). \\
The basis vectors to be associated with
the incoming boundary can be denoted in general as
\begin{equation} \label{ourket}
\left| [ j_1,j_2,...,j_{2N - 1},j_{2N}];\;[\mathbf{k};\,\mathbf{h}];\;0,0\right\rangle, 
\end{equation}
where the last two entries are the quantum numbers $JM $ $\equiv$
$(00)$ for a singlet state of the total angular momentum and 
the first string represents the incoming spin variables (we drop the hat 
on oriented objects whenever not necessary).
The second group of entries denotes particular sets of intermediate angular momentum labels,
arising from binary couplings of the $2N$ primary fields $j$'s,
with a bipartite structure represented by the symbols 
$\mathbf{k} = k_1,k_2,\ldots$ and $\mathbf{h}=h_1,h_2,\ldots$, 
chosen in order to comply with the rules described below. 
Before addressing the latter in general, let us have a look at the simplest 
non trivial case of $N=4$ incoming spin labels. The underlying admissible
combinatorial structures are depicted in Fig.
\ref{RacahOnTrees}.

In this case we have just  two types of basis vectors, 
related by the so--called duality matrix of WZ theory
\begin{equation} \label{dualmatr}
\left| [j_1,j_2,j_3,j_4];[k,k;-];00 \right\rangle=\sum\limits_{k'} \left( \begin{array}{cc}
j_1&j_2\\
j_3&j_4
\end{array} \right. \bigg\vert \left. \begin{array}{c}
k\\
k'
\end{array} \right)_q
\left| [j_1,j_2,j_3,j_4];[k';\tfrac{1}{2}];00 \right\rangle,
\end{equation}
where the vectors on the left hand side do not contain any $\mathbf{h}$--label,
namely $[\mathbf{k},\mathbf{h}]$ $=[k,k;\,-\,]$, while on the right 
we have the combination $[\mathbf{k}',\mathbf{h}']$ $=[k';\tfrac{1}{2}]$.  The array
$(\,::\,|:)_q$ in \eqref{dualmatr}
is the (normalized) $q$--Racah symbol of $U_q(su(2))$ introduced in \eqref{norQracah}
of section 4 (with respect the notation used there, from now on 
we drop the subscript $q$ on vectors).

\begin{figure}[htb]
\begin{center}
\includegraphics[width=6cm]{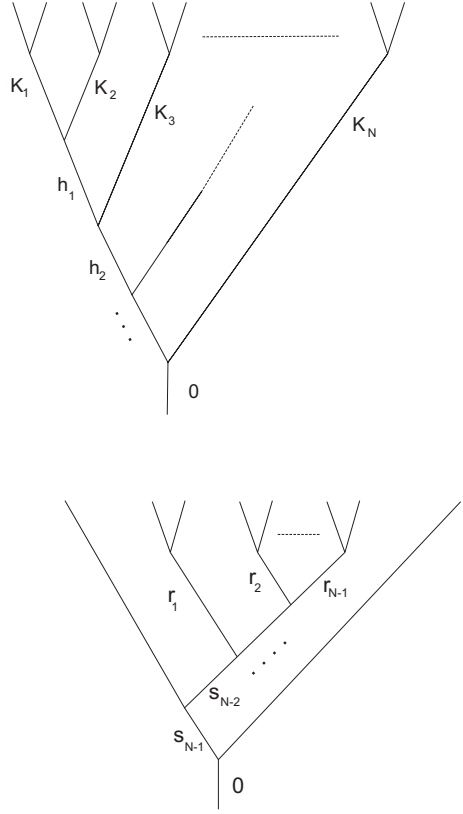}
\end{center}
\caption{Coupling binary trees representing the combinatorics of
the two sets of bases in the case of a generic $N$, see \eqref{evenket} and \eqref{oddket}.}
\label{correlators}
\end{figure}

In the general case of $2N$ incoming spin labels, the two combinatorially
distinct bases
which have to be involved are specializations of the vectors in \eqref{ourket} to
the two configurations depicted in  Fig.
\ref{correlators}.
The extension of the duality  transformation \eqref{dualmatr} 
to the case of an arbitrary (even) number of incoming spins
can be done by resorting to two types of more complicated
arrays, which can be represented as  

\begin{equation}\label{dualityN}
\left( \begin{array}{cccc}
j_1&j_2\\
j_3&j_4\\
\vdots&\vdots\\
j_{2N-5}&j_{2N-4}\\
j_{2N-3}&j_{2N-2}\\
j_{2N-1}&j_{2N}
\end{array} \right. 
\left| 
\left. 
\begin{array}{cc}
k_1&h_1\\
k_2&h_2\\
\vdots&\vdots\\
k_{N-2}&h_{N-2}\\
k_{N-1}&-\\
k_N&-
\end{array} \right. \right)_q,
\left( \begin{array}{cccc}
j_2&j_3\\
j_4&j_5\\
\vdots&\vdots\\
j_{2N-4}&j_{2N-3}\\
j_{2N-2}&j_{2N-1}\\
j_{2N}&j_{1}
\end{array} \right. 
\left| 
\left. 
\begin{array}{cc}
r_1&s_1\\
r_2&s_2\\
\vdots&\vdots\\
r_{N-2}&s_{N-2}\\
r_{N-1}&s_{N-1}\\
-&-
\end{array} \right . \right)_q
\end{equation}
where the matrix indices --to be involved in summations
whenever transformations which generalize \eqref{dualmatr}
are implemented-- are listed in the right hand side of the arrays.\\
As happens in the standard Racah--Wigner setting, 
it can be shown that each of these arrays, 
which represent the $q$--deformed counterparts of $SU(2)$
$3nj$ coefficients, can be decomposed in terms of $q$--Racah 
transformations \eqref{dualmatr}, {\em cfr.} sections 3,4 of \cite{MaRa2} and \cite{Kaul}.

When the outgoing Hilbert space is considered (corresponding to the boundary
$\Sigma_{2}$ endowed with the opposite orientation with respect to $\Sigma_{1}$)
we have to introduce bra--type bases which are dual (and orthonormal) 
with respect to the bases in
\eqref{ourket}. With an obvious choice of notations we set 
\begin{equation} \label{bra}
\left\langle  [ j_1,...,j_{2N} ];\,[\mathbf{k};\mathbf{h}]\,;0,0 \right| \left. 
[ j_1,...,j_{2N}];[\mathbf{k}';\mathbf{h}']\,;0,0 
\right\rangle = \delta_{\mathbf{k},\mathbf{k}'} \delta_{\mathbf{h},\mathbf{h}'},
\end{equation}
where, as before,  $\mathbf{h},\mathbf{k},\mathbf{h}',\mathbf{k}'$ represent 
multi--indices to be associated with 
the admissible configurations  of binary coupled spins and there appear 
multiple Kronecker deltas.

The discussion above was aimed to recognizing the crucial fact that the basis vectors 
\begin{equation} \label{oddket}
\left| [ j_1,j_2,...,j_{2N - 1},j_{2N}];[k_1,...,k_N;h_1,...,h_{N-2}];0,0\right\rangle 
\end{equation}
are eigenfunctions of the odd braiding operators $B_{2l - 1}$, 
while the basis vectors 
\begin{equation} \label{evenket}
\left| [ j_1,j_2,...,j_{2N - 1},j_{2N}];[r_1,...,r_{N-1};s_1,...,s_{N-1}];0,0\right\rangle 
\end{equation}
are eigenfunctions of the even braiding operators $B_{2l}$.\\
The explicit expressions of the eigenvalues, in the odd and even case respectively, read
\begin{equation*}
\lambda_{k_l} (\hat{j}_{2l-1},\hat{j}_{2l})\doteq \lambda _z^{(+)} ( j,j') 
= (-)^{j + j' - z}\, q^{(c_j  + c_{j'})/2 
+ c_{\min (j,j')} - c_z /2} \;\; \text{for} \;\;\epsilon \epsilon'=+1
\end{equation*}
\begin{equation}\label{oddeveigen}  
\lambda_{r_l} (\hat{j}_{2l},\hat{j}_{2l+1})\doteq( \lambda _z^{(-)} ( {j,j'}) )^{ - 1}  
=(-)^{|j - j'| - z} \, q^{|(c_j  - c_{j'}|/2 - c_z /2} \;\; \text{for} \;\;\epsilon \epsilon'=-1.
\end{equation}
Here $l=1,2,\ldots,N-1$, $q$ is the deformation parameter, $z \in \{k_1,k_2,
\ldots ,k_N,$ $r_1,r_2,\ldots,r_{N-1}\}$, $c_z  \equiv z(z + 1)$ 
is the quadratic Casimir for the spin--$z$ representation and
$\epsilon, \epsilon'$ denote the orientation of the strands labelled by
$j$ and $j'$, respectively. Thus 
$
\lambda _z^{(+)}(j,j')
$ 
is the eigenvalue of the matrix which performs a right handed 
half--twist in contiguous strands with the same orientation, while 
$
\lambda _z^{(-)}(j,j')
$ 
is the eigenvalue of the matrix which performs a right handed 
half--twist in strands with opposite orientation. 

The  explicit expression of the formal expectation value given in \eqref{first} above gives, after 
normalization according to the standard conventions ({\em cfr.}  \cite{Kaul}), 
the extended Jones polynomial of the colored link $L$ associated with the braid
\eqref{genbraid}, namely
\begin{equation*} \label{extjones}
V_{j_1j_2 \ldots j_N}\,[L;\,q] =\,\prod_{i = 1}^N \;\,[2j_i  + 1]_q \;\,\times
\end{equation*}
\begin{equation}\label{extjones}
\langle [\hat{l}_1,\hat{l}_1^*,...,\hat{l}_N,\hat{l}_N^*];[0;0];0,0 \mathbf{\lvert} 
\,B\, \bigl( \begin{smallmatrix}
   \hat{j}_1  & \hat{j}_1^*  & \dots & \hat{j}_N  & \hat{j}_N^*   \\
   \hat{l}_1  & \hat{l}_1^*  & \dots & \hat{l}_N  & \hat{l}_N^*  
 \end{smallmatrix} \bigr)
 \mathbf{\rvert} \,[ \hat{j}_1,\hat{j}_1^*,...,\hat{j}_N,\hat{j}_N^*];[0,0];0,0\rangle ,
\end{equation}
where $[2j_i +1]_q$ is the $q$--dimension  of the $U_q(su(2))$ irrep $j_i$ 
defined in section 4.
The operator $B(\,:::\,)$ is expressed 
in terms of (a finite sequence of)  the elementary braiding operators $\{B_{2l-1};B_{2l}\}$, suitably
changed into the current odd (even) basis by acting with  a $q$--duality matrix,
whenever an even (odd) vector of type \eqref{evenket} (\eqref{oddket}, respectively) is encountered. 
According to the expressions
\eqref{dualityN} and \eqref{oddeveigen} for the admissible operations, the running variable
of the polynomial is given by $q=\exp \{-2\pi i/(\ell+2)\}$ for any integer $\ell \geq 3$. 
Moreover, the above expectation value is to be interpreted as a  trace over free spin
labels. This feature derives of course from the geometric construction of the plat 
presentation of the link $L$ outlined at the beginning of this section, since the colored
oriented braid \eqref{genbraid} has to be `closed up' to get the associated link.
More precisely, corresponding  strands must not only match pairwise with the correct orientations
(as is made manifest by our notation), but $j$ and $l$--type labels have to be 
appropriately identified (traced) in pairs, namely the $l$'s in
\eqref{extjones} are not new independent labels.\\
Coming back to the $3$--dimensional picture, such trace procedure amounts
to gluing back the two opposite--oriented boundary $2$--spheres, paying attention 
to the coloring of the punctures, to end up with the same $3$--sphere we started from. 
As already pointed out, the resulting link polynomial, arising as  
vacuum expectation value of the composite Wilson loop
operator \eqref{Wilexpect} in the quantum $SU(2)$ CS theory for $\mathcal{M}^3$ $=S^3$,
is automatically an invariant of regular isotopy ({\em cfr.} the remarks in
 the footnote of section 5.2).

\vspace{12pt}

This long technical discussion about the derivation of the extended Jones polynomial
is nothing but the necessary premise to address the main issue of the present paper, namely
the analysis of the connections among 
the theory of formal languages (section 2) and the spin--network computational
scheme (sections 3 and 4), on the one hand, and braid group and links invariants on the other.
We are going to interpret the results established so far  in terms of 
`processing of words',  written in the alphabet given by the generators of the braid 
group, on a quantum automaton in such a way 
that the expectation value associated with the `evolution' of the automaton is precisely 
the extended Jones polynomial. 
The quantum automaton in question will in turn 
correspond to a path in the $q$--braided spin--network computational graph. 

In order to comply with the requirements for a finite--states quantum   
automaton described in section 2, we have to provide 
explicitly  
the 5-tuple $(Q,\Sigma,\delta,\mathbf{q}_0,F)$,
where $F$ represents a set of acceptable final states. 
Now $Q$ is a finite set of states
belonging to the Hilbert spaces of the tensor algebra of
$U_q(su(2))$ described in section 5.1, whose combinatorial content was depicted in
Fig. \ref{correlators}. 
Labels of $N$ irreps in this quantum
group are associated with 
the strands of the plat presentation of the link $L$, and $\mathbf{B}_{2N}$ is
the braid group to be selected. $\Sigma$ is the alphabet made of
the $2N-1$ generators of $\mathbf{B}_{2N}$: 
each generator (and its inverse) represents a letter of the  alphabet, and words are written as 
composition of these elementary braids.  The function $\delta$ denotes a set of unitary matrices defining 
the transition rules and there is of course one 
matrix for each letter of the alphabet since we are linearly
representing the braid group in the tensor algebra $U_q(su(2))$. 

As initial state $\mathbf{q}_0$ we pick up one particular
binary--coupled state, namely 
\begin{equation} \label{ourinput}
\left| \left[ (j_1,j_2),...,(j_{2N-1},j_{2N})\right];[0;0];0,0\right\rangle ,
\end{equation}
where, with respect to the generic expression for an odd basis vector given in \eqref{oddket},
we choose $[\mathbf{k};\mathbf{h}]$ $=[0;0]$, namely we select a `multi--singlet'
intermediate state.

The set of final states $F$ is constrained by the topological properties of
the plat presentation, namely final states may differ from 
${\bf q}_0$ by a permutation on the string $(j_1 j_2 \ldots j_{2N})$.
Thus we can actually build up a family of $N!$ automata out of one
initial state ${\bf q}_0$.
By acting with the symmetric group on the binary parenthesization structure of \eqref{ourinput}
we may get, for instance, the singlet final state 
\begin{equation} \label{ouroutput}
\left|\left[ (((((j_3,j_2),(j_1,j_4)),(j_{2N},j_6)),...),(j_{2N-1},j_5))\right];[0,0];0,0 \right\rangle.
\end{equation} 

\noindent The unitary transition rules codified in the set $\delta$ are:
\begin{itemize}
	\item if the automaton is in an even (odd) state and it reads an even (odd) braid generator, 
then the system evolves with the R--matrix associated to the proper braid generator, $B_{2l-1}$ or
$B_{2l}$, see \eqref{oddket}, \eqref{evenket} and \eqref{oddeveigen};
	\item if the automaton is in an odd (even) state and it reads an even (odd) braid generator, then the system 
evolves with the proper duality transformation (see \eqref{dualityN}) to update the actual state
into the  configuration consistent with the parity of the given braid generator.
As pointed out before, such a transformation can be splitted into a finite sequence of 
elementary duality ($q$--Racah) transformations of the type \eqref{dualmatr}. 
\end{itemize}

\noindent Once a final state ${\bf q}_{f}$ has been 
selected (the right permutation
can be singled out in a fast way even by a classical machine) the evaluation of
the polynomial \eqref{extjones} is carried out by the automaton in
a number of steps  linear in 
the length $|B|$ of the `word' $B$. Since $|B|$ is the sum
of the numbers 
of elementary braiding operators
and $q$--duality transformations entering the explicit  expression of
$B$, the length of the word is bounded from above by a linear function of
the number of crossings of the plat presentation of the associated link  $L$. 
On the other hand, in the worst case we have to perform one duality
transformation \eqref{dualityN} before applying each elementary braiding
operator $B_i \in \{B_{2l-1}, B_{2l} \}$. This happens, for instance,
in the evaluation of the Jones polynomial of the trefoil knot illustrated in the Appendix.
This latter remark lead us to conclude that the time complexity function
(more precisely, the number of computational steps in our automaton calculation)
equals the length of $B$ and is bounded according to
\begin{equation}\label{complex1}
|B|\;\leq \;\mathfrak{c}(N)\,\kappa (L), 
\end{equation}
where $\mathfrak{c}(N)$ is a positive number depending on the 
number of generators of the braid group $\mathbf{B}_{2N}$
and $\kappa (L)$ is the number of crossings in the plat presentation of the link $L$.\\
We can estimate $\mathfrak{c}(N)$ by observing that any such automaton 
can be uniquely associated with a 
particular path $\mathcal{P}$ in the $q$--braided spin--network computational space 
$(\mathfrak{G}_n (V, E)\, \times\, \mathbb{C}^{2J+1})\,\times \,
\mathbb{Z}_2$ for $n=2N-1$. In particular, the maximum number of elementary
$q$--Racah transforms entering 
a duality matrix of  type \eqref{dualityN} must coincide with the number
of Racah transforms entering into one (classical) $3nj$ symbol since the combinatorics
of such operations is manifestly the same. Hence we may exploit results from graph 
theory which tell us that
the Rotation graph $\mathfrak{\tilde{G}}_n (\tilde{V}, \tilde{E})$ 
--obtained from $\mathfrak{G}_n (V, E)$
by ignoring twists (or braidings)-- has a diameter of the order $n \, \ln n$  
(the diameter is defined as the
maximum over the set of distances between pairs of vertices, where the distance is the minimum
number of edges connecting two given vertices) ({\em cfr.} \cite{Belgi} and appendix A of \cite{MaRa2}
for a complete discussion of the spin--network combinatorics). 
Clearly the `distance' between the current basis and the eigenbasis with the
right parity cannot exceed the above maximum distance, and consequently 
the factor $\mathfrak{c}(N)$ in \eqref{complex1} grows as 
\begin{equation}\label{complex2}
\mathfrak{c}(N)\,\sim \,(2N-1)\, \ln \,(2N-1), 
\end{equation}
namely polynomially in the number of strands of the link.

A deeper connection with the $q$--braided spin--network computational 
scheme comes out however
 when we recognize that the expectation value \eqref{extjones} 
representing the extended Jones
polynomial is not
only the quantum transition amplitude of a finite states--automaton, as pointed out before,
but complies also with the 
expectation value \eqref{snexp2} to be 
associated with a path $\mathcal{P}$ in the $q$--version of the spin--network computational space.
In this new perspective,
what we are really doing is to `encode' the combinatorial structure underlying quantum $SU(2)$
Chern--Simons field theory (and the associated  WZ boundary theory)
at some fixed level $\ell$ into the 
abstract $q$--braided $SU(2)$--spin--network for   $q= \exp \{-2\pi/(\ell +2)\}$. 
This does not mean, of course, 
that we have set up a
quantum algorithm for the extended Jones polynomial in the strict sense, 
since the encoding map could not be `efficiently' represented (nor efficiently approximated)
with respect to standard models of computation
(Boolean circuits, Turing machines). We provide, however, a 
 quantum system whose
evolution can be controlled in such a way that its probability amplitudes
give the desired link polynomials.

The crucial
issue of constructing a {\em bona fide} quantum
algorithm is under investigation. It will require in particular: 
  {\bf i}) (efficient) encoding schemes for binary coupled states; {\bf ii})
(efficient) algorithms to evaluate (or approximate) the
basic operations, namely the Racah transform and its associated $6j$--symbol
for arbitrary entries on the one hand, and the elementary braiding operators
on the other.
  
As a final remark we notice that the field theoretical approach gives us automatically 
the rate of growth of the absolute value of extended Jones polynomial with respect
to the Chern--Simons coupling constant $k=\ell+2$. The absolute value of the Reshetikhin--Turaev  \cite{ReTu}
quantum invariants of $3$--manifolds, $|Z^G_k (\mathcal{M}^3)|$ (which are linear combinations
of colored polynomials associated with surgery framed links $\subset \mathcal{M}^3$) are
estimated to grow as $\mathcal{O} (k^d)$, where the exponent $d$ is bounded from above by some
simple function (depending on the gauge group $G$) of the Heegaard genus of the manifold
({\em cfr.} \cite{Oht}, Ch. 7).

\section*{Appendix}

\begin{figure}[htbp]
\begin{center}
\includegraphics[width=3cm]{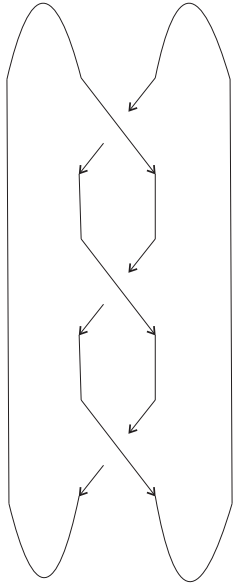}
\end{center}
\caption{Plat presentation of the oriented trefoil knot.}
\label{trefoilEx}
\end{figure}

\noindent A simple application of the procedure described in section 6 for the evaluation of 
quantum link invariants is the explicit computation of the Jones polynomial for the
plat presentation of the  trefoil 
knot $K_{\text{tref}}$ depicted in Fig. \ref{trefoilEx}. The four strands are labelled by a same
$j_1$, together with its opposite $j^*_1$, from left to right. 

Accordingly, the initial and final
states to be associated with the quantum automaton are
\begin{equation} \label{intrefoil}
\left|\left[(((j_1,j_1^*),(j_1^*,j_1))) \right];[0,0;-];0,0 \right\rangle
\end{equation}
and 
\begin{equation} \label{outtrefoil}
\left|\left[(((j_1^*,j_1),(j_1,j_1^*))) \right];[0,0;-];0,0 \right\rangle
\end{equation}
respectively, and they comply  with the prescription of being odd vectors, see
\eqref{ourinput}. From the picture we easily recognize that the operator to be 
employed is $B_{\text{\,tref}}=(B_2)^3$.
Since we are interested in the evaluation of the Jones polynomial we set 
from now on $j_1, j_1^*\equiv \tfrac{1}{2}$. Moreover, in order to 
apply the even braiding operator $B_2$ we have to perform preliminarly
a duality transformation \eqref{dualmatr} on the odd vector \eqref{intrefoil}
\begin{equation} \label{RacahTref}
\left| [\tfrac{1}{2},\tfrac{1}{2},\tfrac{1}{2},\tfrac{1}{2}];[0;-];00 \right\rangle
=\sum\limits_{l=0}^1 
\left( \begin{array}{cc}
\tfrac{1}{2}&\tfrac{1}{2}\\
\tfrac{1}{2}&\tfrac{1}{2}
\end{array} \right. \bigg\vert \left. \begin{array}{c}
l\\
0
\end{array} \right)_q
\left| [\tfrac{1}{2},\tfrac{1}{2},\tfrac{1}{2},\tfrac{1}{2}];[l;-];00 \right\rangle,
\end{equation}
which converts the initial state of the automaton into eigenvectors of 
the braid generator $B_2$. The application of $B_{\text{\,tref}}$ gives 
\begin{equation} \label{EigenEquation}
(B_2)^3\left| [\tfrac{1}{2},\tfrac{1}{2},\tfrac{1}{2},\tfrac{1}{2}];[0;-];00 
\right\rangle=\sum\limits_{l=0}^1 (\lambda_l^{(+)})^3
\left( \begin{array}{cc}
\tfrac{1}{2}&\tfrac{1}{2}\\
\tfrac{1}{2}&\tfrac{1}{2}\end{array} \right. \bigg\vert \left. \begin{array}{c}
l\\
0
\end{array} \right)_q
\left| [\tfrac{1}{2},\tfrac{1}{2},\tfrac{1}{2},\tfrac{1}{2}];[l;-];00 \right\rangle,
\end{equation}
where there appears the cube of the eigenvalue $\lambda_l^{(+)}$ defined in \eqref{oddeveigen}.
According to the expression of the extended Jones polynomial
given in \eqref{extjones} and taking into account \eqref{outtrefoil}, we get
\begin{equation} \label{ExpValueTrefoil}
V_{j=\tfrac{1}{2}}\,(K_{\text{tref}}\,;q)\,=
[2]_q\;\langle\left[\tfrac{1}{2},\tfrac{1}{2},\tfrac{1}{2},\tfrac{1}{2}];[0;-];00\right|\,(B_2)^3
\left| [\tfrac{1}{2},\tfrac{1}{2},\tfrac{1}{2},\tfrac{1}{2}];[0;-];00 \right\rangle,
\end{equation}
which, by using the orthogonality relations of the duality matrices (KAUL), amounts to
\begin{equation} \label{EvaluationTrefoil}
V_{j=\tfrac{1}{2}}\,(K_{\text{tref}}\,;q)\,=[2]_q\sum\limits_{l=0}^1 \lambda_l^{(+)^3}=\frac{-1+q+q^3}{q^4}
\end{equation}
as required.

\end{document}